\documentclass[aip,twocolumn,reprint]{revtex4-2}
\usepackage{float}
\usepackage{graphicx}
\usepackage{bm}
\usepackage{hyperref}
\hypersetup{
    colorlinks=true,
    linkcolor=blue,
    filecolor=magenta,      
    urlcolor=black,
}
\usepackage{color}
\usepackage{fancyhdr}
\usepackage[normalem]{ulem}
\usepackage{physics}
\usepackage[dvipsnames]{xcolor}
\usepackage{ulem}
\begin{document}
\title{LETTER\\ On putative incommensurate states of a finite Frenkel-Kontorova chain}
%
\author{W.\,Quapp}
\email{quapp@math.uni-leipzig.de \\ } 
\homepage{http://www.mathematik.uni-leipzig.de/MI/quapp}
\affiliation{Universit\"{a}t Leipzig,
Mathematisches Institut, PF 100920,  
D-04009 Leipzig, Germany (ORCID: 0000-0002-0366-1408)}
\author{J.\,M.\,Bofill}
\email{jmbofill@ub.edu} 
\affiliation{Departament de Qu\'{i}mica Inorg\`{a}nica i Org\`{a}nica,
Secci\'{o} de Qu\'{i}mica Org\`{a}nica, and
Institut de Qu\'{i}mica Te\`{o}rica i Computacional, (IQTCUB),
Universitat de Barcelona, Mart\'{i} i Franqu\`{e}s 1, 08028 Barcelona, Spain 
(ORCID: 0000-0002-0974-4618)}
\date{ preprint \today }
%
\begin{abstract}
We propose a new but simpler explanation of the phases of 
a Frenkel-Kontorova chain of atoms, and we demonstrate it by examples. 
Combined with this, we present a criticism of the theory of 
so-called commensurate and incommensurate states, 
especially for finite chains. 
We reject the putative observation 
of an Aubry-phase transition in a finite chain. 
\end{abstract}
%
\maketitle 
\section{Introduction}
This letter is devoted to the aim of understanding what happens 
inside the finite Frenkel-Kontorova (FK)   
chain~\cite{kont38I,kont38II,QuBo18MolPhys}, 
if the two characteristic  ratios $a/b$ and $V_o/k$ change.
We select ref.~\onlinecite{mun20} to discuss some general problems of 
the understanding of the FK model, with the 
special example of this paper.
For shortness we only treat the springlike potential of ref.~\onlinecite{mun20}. 

The corresponding model of the potential energy 
of a linear FK chain $\textbf{x}=(x_1,...,x_N)$ with 
$x_i<x_{i+1} $ for all atoms, and with 
atoms of equal mass at points $x_i$ is ~\cite{mun20}
\begin{equation}
U(\textbf{x}) =  V_o\, \sum_{i=1}^N(1- cos(\frac{2\pi}{b}\,x_i)) +
 \frac{k}{2} \sum_{i=1}^{N-1}V_{int} (x_{i+1}-x_i) \ .
\label{pot}  
\end{equation}
$V_o$ describes the strongness of the substrate but $k$ describes 
the strongness of the springlike forces between neighboring atoms.
Used are $V_o=0.02 ,  k/k_B=783.6$ (without units~\cite{mun20}).
Parameter $b$ is the periodicity of the substrate. It is used with a  
variable length around the spring distance of the $V_{int}$ potential, $a=2.4$.
This interatomic springlike potential function is
\begin{equation}
V_{int}(r)= \frac{-1}{1+c_o(r-r_o)^2}+\frac{1}{k} Exp(-c_2(r-r_o')) \ .
\label{spring}
\end{equation}
The following parameters of the function are used~\cite{mun20}
$c_o$=12.9, $r_o$=2.4, $c_2$=263.0, $r_o'$=1.8.   

\begin{figure}  
\includegraphics[scale=0.875]{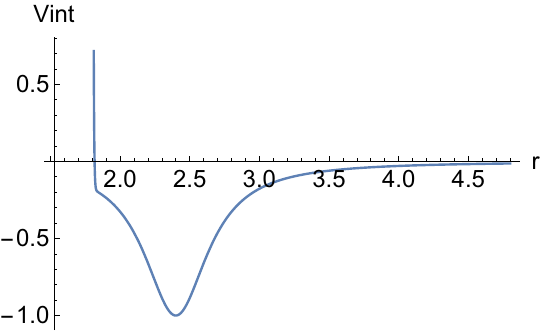}
\caption{The springlike potential function of Eq.(\ref{spring}).}
\label{fig1}
\end{figure}

The graph of function (2) is shown in Fig.\ref{fig1}.
The minimum is found to be at $a=r_{min}=2.4$ units, 
in contrast to ref.~\onlinecite{mun20} with 2.44 units.
We used the Mathematica program, version 13.0, for the calculations, as well as for the 
 Figures.

The springlike potential is of special interest because it is unsymmetrical for      
stretching against compression. 
However for the calculation of minima this fact alone does not
play a role.
The boundary conditions for $x_1$ and $x_N$ are free.
%
Parameter $a=r_{min}$ is the distance of two atoms if the parameter $V_o$ 
is  zero. It is hold fixed throughout the letter. 
%
However, in the chain with the substrate potential,
$V_o\ne 0$, usually the average distance, $\tilde{a}$, changes to
\begin{equation}
 \tilde{a} = \frac{x_N-x_1}{N-1} \ ,
\end{equation}
because the chain will find a minimum form in the substrate\cite{stoy85},
compare Fig.2.
We have to emphasize that $ \tilde{a}$ is the average distance 
in a minimum structure.

In our calculations on the relation of the influence of the two parts of Eq.(1) 
we only change the parameters $b, \ V_o$ of the substrate potential. 
The spring potential and $k$ are fixed.  

\begin{figure}[h]  
\includegraphics[scale=0.75]{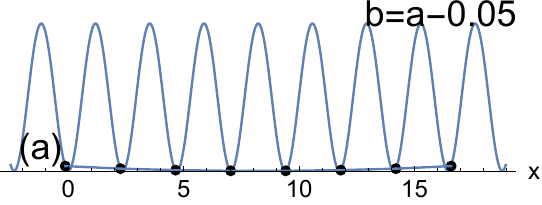}
%
\includegraphics[scale=0.75]{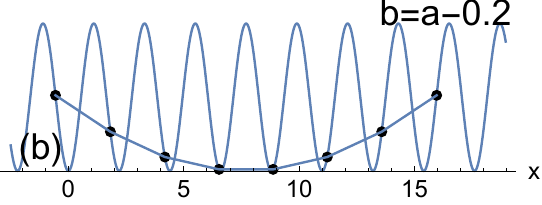}
%
\includegraphics[scale=0.75]{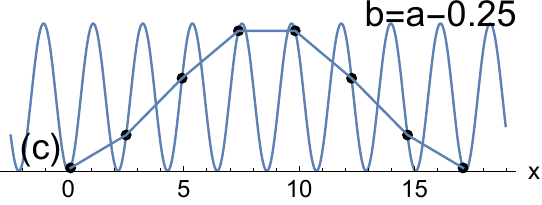}
%
\includegraphics[scale=0.75]{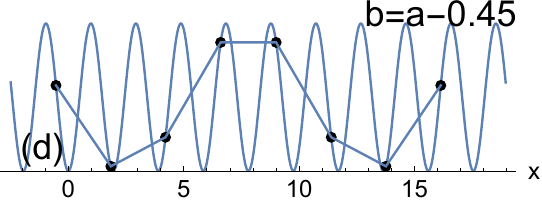}
%
\includegraphics[scale=0.75]{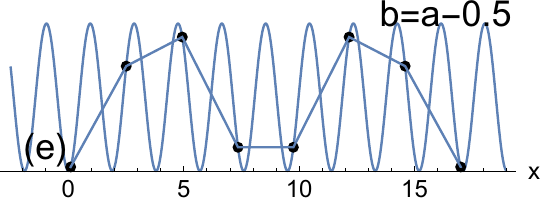}
%
\caption{Ground state of FK chains of different $b$ values.
Only the substrate potential is shown, however, 
the fixed springlike force is not shown. 
To lead the eye, we moved the atoms up to the potential line.
Note the different length of the substrate throughs.
In (c) and in (e) new kinks emerge.
The outer atoms show impressively the effect of the free boundaries.}
\label{figX}
\end{figure}

\section{COMMENSURATE MINIMA}
Our remark concerns the discussion of commensurate or incommensurate
states of the chain. 
Note that this letter is similar to another one~\cite{QuBo22StatMech}. 
Thus, the problem may be widespread distributed in the FK community.

We start with the chain at the potential energy (1).  
 We determine the minimizer 
of the chain in the combined two potentials (1), by the substrate and by the 
springlike forces. 
The definition of a 'commensurate' chain in ref.~\onlinecite{mun20} is useless, 
because it is empty if $Q>N$ 
\begin{equation}
x_{Q+i}= x_i + R\,b 
\label{empty}
\end{equation}
for integers $Q,\,R$.
Then $x_{Q+i}$ would be outside the chain.
For example it is the case for $N=50$ and $R/Q=52/51$.  
Note that cases $Q>N$ will come far before an irrational relation 
of $\tilde{a}/b$.
Of course, for nearly all cases with  $a \ne b$ we will find some atoms at $x_i$ 
outside the bottom of their corresponding through.

We have recalculated a part of a similar curve like in Fig.4, part (a-3)~\cite{mun20}. 
A result is shown in our Fig.\,2 by some special chains belonging to the case
$V_0=0.9\,k$.
Calculations are executed for steps $b=a-s*0.05$ for the step numbers
$s=0,..,10$ in the $[a-0.5,a]$ interval for $b$.
The springlike potential with parameter $a$ is not changed.
In every case, we 'sort' the constantly equal chain 
into different substrate throughs. 
And there it changes, see Fig.\,\ref{figX}.

To see what happens in the FK chain, we  only used $N=8$ atoms 
(and not 50 like it is done  in ref.~\onlinecite{mun20}).
We always start the minimization of the chain with $x_1=0$ 
and $x_i=a\,(i-1)$ for $i=2,...,N$.
We get steps like in Fig.4 of ref.~\onlinecite{mun20}.
Of course, 
only rational numbers are used for the steps.
The step near the line 1 is not constant; 
it slowly increases to the right hand side.
%

We conclude that the statement that there are
 ``small intervals of zero slope'' \cite{mun20,chai95} is not correct.
The increase comes from a small ascent of the $\approx N/2$ left atoms 
to the left walls of their corresponding substrate potential throughs, 
and of the $\approx N/2$ right  atoms   
to the right walls of their corresponding substrate throughs, 
what increases $\tilde{a}$. 
This implies that this fact does not have relations with commensurate, 
or incommensurate numbers.

We explain the steps in Fig.2. 
It is $\tilde{a}=$ 2.39, 2.36, 2.25, 2.38, and 2.38 correspondingly, and 
$\tilde{a}/b$ is then 1.02, 1.07, 1.13, 1.22 and 1.25.
Thus we really get consecutive states on the staircase like in Fig.4 
of ref.~\onlinecite{mun20}.

The step from part (a) to (b) for a shorter $b$, 
thus a larger $a/b$, shows a climbing up of the outer atoms to 
their left walls (the $N/2$ left ones) or to 
the right hand side walls (the right $N/2$ atoms) .

Between (b) and (c) happens a 'phase' jump, see below a definition of a phase.
Now the chain occupies one through more of the substrate, not $N=8$ however 9 waves.
This is possible because the central through, say with number 5, is now empty. 
The chain has a kink. 
Note that all structures in Fig.2 are minima.  

Part (d) is again a stretching of the chain, but $a/b$ correctly increases.

A next 'phase' jump happens between (d) and (e). Again a new kink emerges. 
Now two kinks concern the throughs 3 and 8 of the substrate, 
which are bridged by the kinks. 

The key is that one or two atoms, $x_5$ in (c), or $x_3$ and $x_8$ in part (e), 
jump at the end of the increasing movement over a peak of the site-up potential.
{
Note that the new kinks are not restricted to the center of the chain, 
as it has been claimed in a former work~\cite{mikh02}.
The pattern of new kinks
}
repeats for higher values of $a/b$ again and again at every step. 
So  the jumps emerge in Fig.4 of ref.~\onlinecite{mun20}. Thus, the given
jumps are correct.

The pure minima of the FK-chain with $N=8$ atoms disclose
3 'phases' in the region $1\le a/b\le 1.265$.
The minimization of Mathematica leads to the same
phase for certain intervals of $a/b$. 
The character of the used number $b$ rational or irrational, is irrelevant. 
What counts is the length of the chain, given by the $a$, and the (integer) number of
throughs of the substrate, which the chain occupies. Any distinction into 
commensurate or incommensurate is nonsense here. 

From Cantor's proof that the real numbers are uncountable but the
rational numbers are countable, 
it follows that almost all real numbers are
irrational. But the rational numbers  are dense as well as the irrational ones.
There is no gap in both kinds of numbers
which indicates the steps in Fig.4 of ref.~\cite{mun20}

The statement of ref.~\onlinecite{mun20} that 
"the ratio $a/b$ is highly irrational and quite far from a rational value"
is not correct. There is no smallest distance from every irrational number to
 any "neighboring" rational one. If $z$ is irrational, and if
 such a distance existed then we coud take the 
"left" rational number $z_1$ and the "right" rational number $z_2$ and 
build the rational number 
$
(z_1+z_2)/2
$
which is nearer to the initial $z$ number.

We conclude that relating the transition of different phases of the 
FK-chain with the transition from a rational to an irrational ratio
of $\tilde{a}/b$ is mathematically and also physically not correct.
And we wonder how the authors of ref.~\onlinecite{mun20} have decided 
the different kind of results in their Fig.4  to be
 'rational' or 'irrational' steps?
We have to assume that they used a computer for their calculation.
Every numerical calculation on a computer goes on with a (restricted) 
range of rational numbers. Irrational numbers are not representable.
We cannot imagine how they use an irrational $b$, and we cannot imagine
how they know that the result $\tilde{a}$ is 'irrational'.

The adsorption of the chain by the substrate acts in the way that for
every ratio $a/b$ a rational $\tilde{a}$ emerges (to every given 
exactness of the solution) for a minimum structure, if $N$ is finite.
The latter was our assumption because in physics only finite chains exist.
Every limit $N\rightarrow \infty$  is a mathematical abstraction.
Nowhere in ref.~\onlinecite{mun20}  
we could find a treatment of such a limit process.

Thus one can assume that to every constellation of the parameters  
 $a/b$, $V_o/k$, and $N$, exits at least one minimizer, 
a structure of the FK chain in a minimum of the combined potential (1).
We now explicitely define different 'phases' of the chain.
\section{Definition of a phase}
If an equilibrium structure of the FK chain occupies $L$ throughs of 
the substrate then it belongs to phase $L$.
There is a certain set of parameters for different ratios 
of $a/b$, $V_o/k$, and $N$, which lead the the same phase.
\\
A single phase transition is correspondingly a change of an 
equilibrium structure of the FK-chain to $L+1$ or $L-1$ throughs.
(We have outlined above that the ``commensurate-definition'' 
 by Eq.\,(\ref{empty})  may be empty.)
For a phase transition the whole chain  contracts, 
or expands, so that one or more atoms of the chain climb 
over their current peaks of the substrate potential. 
As a result, the chain uses less, or more throughs. This is connected 
by a jump in the average $\tilde{a}$, see Fig.\,\ref{figX}.
The question circles around the count of integers.

Besides the ratio of a/b the balance of $k$ and $V_o$  also
plays a role. In all cases, the chains like in Fig.\,\ref{figX}  
are regular structures, no kind 
of incommensurability emerges at any $a/b$ as it is claimed 
in ref.~\onlinecite{mun20}.
\\

We still treat the case $N\rightarrow \infty $. 
In ref.\,~\cite{mun20} is treated 
the finite case only, for $N=50$.
We find that the larger $N$ is, the 'earlier' a small change in 
parameter $b$ leads to a change of the number of occupied throughs.
Because the unperturbed chain ends at $x_N=(N-1)\,a$.
If then $b$ is so much smaller that holds
\[
  N \  b \ < (N-1)\ a 
\]
then the minimization leads to a kink. It means a step in the 'staircase'.
However, in the condition
\begin{equation}
 a > \frac{N}{N-1} \ b
\label{bbb}
\end{equation} 
the factor at $b$ converges to 1 in $N$ for every finite 
 $V_o$ and $k$.
However, for larger and larger $N$, thus the limit treatment,
formula (\ref{bbb}) leads to the decrease of the step length of 
the staircase of Fig.\,4  in ref.~\onlinecite{mun20} to zero.
The staircase of Fig.\,4 in ref.~\onlinecite{mun20} degrades to a straight line.
Any kind of  ``devils staircase'' \cite{mun20,chai95} disappears in the limit.
\section{Phase Changes}
A next section concerns the putative Aubry-phase transition 
demonstrated by Fig.\,6 in ref.\cite{mun20}
The first eigenvalue of the second derivatives of the 
potential of the chain is the frequency for a 
collective movement of the chain.
For an unpinned chain it has to be zero, thus the potential has to be flat.  
However, here the first eigenvalue is greater than zero, for the
blue points, as well as for the red points of Fig.\,6 in ref.\cite{mun20}. 
Thus the potential is a minimum, in both cases, and the chain is pinned. 
Of course, the pinning is small if the $V_o$ is small.
The chain can collectively vibrate, however it cannot move.
The step in the higher eigenvalues under the red points concerns inner 
vibrations of the chain.
This has nothing to do with pinned or unpinned states. 
Unpinned versions of a ground state of
a finite FK chain are not correctly calculated. 
(If $V_o$ is not zero. And if $V_o$ is 
zero then we do not have the FK model, at all.)
\section{Discussion}
We ask for the fascination of irrationals for workers in the field of the FK model.
The early source may be the papers of 
S.Aubry~\cite{aubry83,peyr83,aubry83fr,aubryDae83,axel87,joha08}.
Our main contradiction to these works is the use of an infinite chain. 
For any 'limit'-process which one needs to treat  such a chain, 
the correct way would be 
to study a finite chain with $N$ particles, to determine its equilibrium 
structure, and then to increase step by step the number, $N$, of the chain, 
more and more up to a limit. 
However, such a treatment is missing in the papers of S.Aubry. 

Note, an actual infinity is not possible, as well as an infinite long chain.
Such a construct does not exist in reality. 
In Mathematics only one studies the possibility
of infinite rows, for example, and their convergence, or divergence.
There are strong rules for the handling of the 'infinity'.

Especially, we miss the treatment of the boundary conditions (BCs), 
over which the distance of the chains spring, $a=r_{min}$, (see Fig.\,1) 
comes into play. 
Without the BCs this  parameter of the chain disappears in many FK studies. 
This is questionable.  
If one has free BCs 
then we cannot start with a fixed `left' BC because the minimization 
will result finally in a probably other BC. 
So all the steps of the twist map start in a nebula.
The development by S.Aubry~\cite{aubry92}, 
and others~\cite{black91,bich06,vano2020} which use the twist map, 
ignores that the chain will find at the boundary another minimum structure, 
in comparison to any twist map result\cite{QuBo18MolPhys}.
If there are assumed fixed BCs,  for example, and one starts 'left' with 
the left BC, and one uses all the steps throughout the 
twist map then usually the result at the 'right' end of 
the chain will not fit the right BC.
(However, nobody can put BCs at infinity, 
and nobody can really start at minus infinite, at all.)

We still note that some workers reduce their 
treatment~\cite{shar84,bahu01,mikh01,garc07,wang2011,yjma14,byl16,babu16,novak17} 
(to name a few) to finite chains, 
what is quite correct, however, they continue using the erroneous 
contrast of rational to irrational numbers in the finite FK model. 

In a positive contrast, there is a treatment~\cite{thom22} 
which sorts the FK chain in
a 'commensurable' kind into the site-up potential.
\section{Conclusion}
This letter discusses the widespread theory of so-called commensurate (C)
versus incommensurate (IC) phases of the FK model\cite{mun20,chai95}.
We have seen that the C-phases  
are not specially ordered phases in the sense that
all atoms are locked-in to the minima of the substrate.
In the putative IC-phases we find no broken regular arrangement of the 
atoms of the chain, either. 
What makes steps in the average distance $\tilde{a}$ of the chain is the 
possibility that the chain contracts or stretches over different periods $L$ 
of the substrate.

In a 2D or 3D crystal lattice a long-range periodic order 
with an irrational ratio of the periodicities\cite{chai95,perv2000} can exist. 
Its description as an IC crystal is in order. 
However, the $\tilde{a}$ in the FK model is not the description of 
a fixed lattice. It is the result of the balance of the four different
parameters, $a,\ b,\ V_o,\ k$, and it is only an average value.
The term IC means 'out of proportion'. However, in the FK chain we have to sort 
$N$ atoms into $L$ basins of the substrate, both $N$ and $L$ are integers.
Two integers always form a proportion, 
thus the coined term 'IC' is a wrong term.
One should use the already introduced terms kink and antikink.

The point of $a/b$ where the 'phase' transition occurs, 
thus the chain contracts  or stretches to 
another number of basins of the site-up potential, has nothing to do with 
rational or irrational numbers.

As discussed above, one of the origins of the incorrect 
'C-IC sight' may be due to the theory of S.\,Aubry. 
There is already 
an analysis of this non-appropriete  
 theory~\cite{QuBo18MolPhys,QuBo22StatMech}. 
%
%
%
%
%
\begin{acknowledgments}
We acknowledge the financial support from the Spanish Ministerio de   
Economıa y Competitividad, Project No. PID2019-109518GB-I00, 
and Spanish Structures of Excellence Maria de Maeztu program    
through Grant MDM-2017-0767.
\end{acknowledgments}
\section*{References}
\vspace*{-0.5cm}
%
%
%
%
\end{document}